\documentclass[showpacs,preprintnumbers,amsmath,amssymb,pre,twocolumn]{revtex4}
\usepackage[dvips]{graphicx}
\usepackage{dcolumn}
\usepackage{amsmath}
\usepackage{amssymb}
\usepackage{bm}
\usepackage{xspace}

\newcommand{\atan}{\mathrm{atan}}
\newcommand{\Eqref}[1]{Eq.~\ref{#1}}

\newcommand{\Figref}[1]{Fig.~\ref{#1}}

\newcommand{\bvec}[1]{{\bf #1}}

\begin{document}

\title{Criticality versus $q$ in the $2+1$-dimensional $Z_q$ clock model}
\author{J. Hove}
  \email{Joakim.Hove@phys.ntnu.no}
\author{A. Sudb{\o}}
  \email{Asle.Sudbo@phys.ntnu.no}

  \affiliation{%
     Department of Physics \\
     Norwegian University of Science and Technology, \\
     N-7491 Trondheim, Norway}
\date{\today}


\begin{abstract}
  Using Monte Carlo simulations we have studied the $d=3$ $Z_q$ clock
  model in two different representations, the phase-representation and
  the loop/dumbbell-gas (LDG) representation. We find that for $q \ge
  5$ the critical exponents $\alpha$ and $\nu$ for the specific heat
  and the correlation length, respectively, take on values
  corresponding to the case $q\to \infty$, where $\lim_{q \to \infty}
  Z_q = 3DXY$ model. Hence in terms of critical properties the
  limiting behaviour is reached already at $q=5$.
\end{abstract}
\maketitle

Matter coupled-gauge field theories in $2+1$ dimensions have come
under renewed scrutiny in the context of condensed matter physics in
the past decade, as effective theories of strongly correlated
system\cite{baskaran}.  Concepts such as confinement-deconfinement
transitions associated with the proliferation and recombination of
topological defects of gauge fields, enter for instance in attempts at
providing a theoretical foundation for breakdown of Fermi-liquid
theory in more than one dimension. A large variety of such gauge-field
theories have been proposed, and one model of particular interest is
the compact abelian Higgs
model\cite{baskaran,KNS1,Smiseth:2003,Sudbo:2002}. This model consists
of a compact gauge field coupled minimally to a bosonic scalar field
with the \emph{gauge charge} $q$. In a particular limit the dual of
this model reduces to a loop-gas representation of the global $Z_q$
model\cite{Smiseth:2003,Sudbo:2002}. This identification has been the
motivation for the current work, for a detailed account of the $q$
dependence of the full theory we refer to Ref.
\onlinecite{Smiseth:2003,Sudbo:2002}.

The spin $Z_q$ model is a simple planar-spin model, where the
direction of the spin is parametrized by a phase. This phase is
restricted to the values $2\pi n/q$ with $n \in \mathbb{Z}$, and is
defined by the following action
\begin{equation}
  \label{PhaseZ_q}
  S  = -\beta\sum_{\langle i,j \rangle} \cos \left(\frac{2\pi}{q}\left(n_i - n_j \right)\right).
\end{equation}
The state is specified by the integer variables $n_i \in
\left[0,1,\cdots,q-1\right]$.  Special cases include $q=2$ which is
the Ising model, $q=3$ which is the three state Potts model, and the
limit $q \to \infty$ which corresponds to the XY model. In addition it
is easy to see that for $q=4$ the partition function $Z(2\beta,4) =
Z(\beta,2) \times Z(\beta,2)$. The aim of the current paper has been
to determine how the critical properties interpolate between the well
known Ising ($q=2$) and XY ($q \to \infty$) limits. We have done this
by measuring the exponent combination $(1 + \alpha)/\nu$ as a function
of $q$.

In $d=2$ the model has a quite peculiar phase structure, with an
intermediate \emph{incompletely ordered phase} (IOP), where the system
shows behaviour similar to the critical Kosterlitz Thouless phase.
Upon further cooling, the system will order completely into one of the
$q$ completely ordered states\cite{Elitzur:1979,Norikazu:2002}. In
$d=3$ the $Z_q$ model does not have an IOP, but there are
generalisations of the model which do \cite{Norikazu:2002,Ueno:1991,Ueno:1993}.

A related case is that of a globally $U(1)$ symmetric theory which is
perturbed by a weak crystal field. Using RG and duality arguments it
has been shown that for $q \geq 5$ the crystal field is an irrelevant
perturbation, whereas for $q \leq 4$ the XY fixed point is rendered
unstable\cite{Jose:1977}. 

It is important to emphasise that we have focused on the properties of
the $Z_q$ model \emph{at} the critical point. For $T< T_c$ the
discrete nature of the model will always be apparent. A beautiful RG
study of the $Z_6$ model shows how the couplings of the model flow
towards a fixed point which is ultimately different from the 3DXY
fixed point in the $T \to 0$ limit\cite{Blankschtein:1984,Oshikawa:2000}

\Eqref{PhaseZ_q} is strahtforwardly reformulated as a model of an
interacting ensemble olinks which either form closed loops or
originate and termine at point charges. We start with the partition
function

\begin{equation}
  \label{Z_qPart}
  Z(\beta,q) = \sum_{\{n_i\}} \exp\left[\beta \sum_{i} \left( \sum_{\hat{\mu}} \cos \left(\frac{2\pi}{q}\Delta_{\hat{\mu}}n_i \right) \right)\right].
\end{equation}
The first step is to replace the cosine with a quadratic potential,
this is the Villain approximation\cite{Villain:1975}. Next, we
promote the integers $n_i$  to real-valued phase variables $\theta_i$, at the
expense of introducing an auxiliary integer field $Q$, which through
the Poisson summation formula\cite{Kleinert:1989:book} restricts the
$\theta_i$ variables to the discrete values allowed by original
theory. The resulting partition function is then given by
\begin{widetext}
\begin{equation}
  \label{Z_qVillain}
  Z_V(\beta,q) = \Xi[\beta] \int D\theta \sum_{\{\bvec{k},Q\}} \exp \left[-\sum_{i} \left( \frac{\beta_V}{2} \left( \Delta \theta_i - 2\pi \bvec{k} \right)^2 + 
i q\theta Q \right)\right]. 
\end{equation}
\end{widetext}
In \Eqref{Z_qVillain}, $\{\bvec{k}\}$ is an integer \emph{link field}
living on the links of the \emph{original} lattice, and $\{Q\}$ is a
scalar field living on the \emph{sites} of the same lattice. The
prefactor $\Xi[\beta]$ and effective coupling $\beta_V =
\beta_V(\beta)$ must be retained to get results which agree with
\Eqref{Z_qPart} on a \emph{quantitative}
level\cite{Kleinert:1989:book}, however they do \emph{not} affect the
critical properties and from now on we will assume $\beta_V = \beta$,
$\Xi[\beta] = 1$, and omit the $V$ index on the partition function.

In \Eqref{Z_qVillain}, the $Q$-field explicitly accounts for the
discrete nature of the $Z_q$ model. Setting $Q \equiv 0$, we 
recover the Villain representation of the XY model. Due to this
similarity, the remaining analysis follows well known steps
\cite{Stone:1978}, which we briefly include for 
completeness. A Hubbard-Stratonovich decoupling of the 
quadratic expression in \Eqref{Z_qVillain} is performed
by introducing an auxiliary field $\bvec{v}$, thus bringing
the partition function onto the form
\begin{widetext}
  \begin{equation}
    \label{Z_qHS}
    Z(\beta,q) = \int D\bvec{v} D\theta \sum_{\{\bvec{k},Q\}} \exp 
    \left[-\sum_{i} \left( \frac{1}{2\beta} \bvec{v}^2 +  i\bvec{v} \cdot \left( \Delta \theta_i - 2\pi \bvec{k} \right) + i q\theta Q \right)\right] .
  \end{equation}
\end{widetext}
In \Eqref{Z_qHS} the $\left\{\bvec{k}\right\}$ summation can be
performed, thereby restricting the velocity field $\bvec{v}$ to
integer values denoted by $\bvec{l}$. In the term coupling $\Delta
\theta$ and $\bvec{l}$, a partial integration can be performed, such
that $\theta$ only appears in the combination $i\theta\left( \Delta
  \bvec{l} - qQ \right)$, from this we get the constraint
\begin{equation}
  \label{constraint}
  \left( \Delta \bvec{l} - qQ\right) = 0.
\end{equation}

At this stage the transformation to a loop gas is complete, and the
partition function is given by
\begin{equation}
  \label{ZLoopGas}
  Z(\beta,q) = \sum_{\{\bvec{l},Q\}} \delta_{\nabla \bvec{l},qQ} \exp \left[ \frac{-1}{2\beta} \sum_{i} \bvec{l}^2 \right].
\end{equation}

\begin{figure}[htbp]
  \centerline{\scalebox{0.5}{\rotatebox{0.0}{\includegraphics{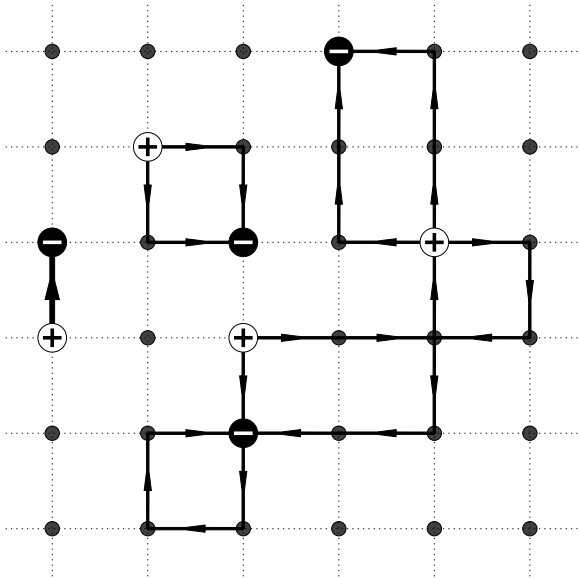}}}}
\caption{\label{Z2Config}A typical LDG configuration for the $q=2$
  (Ising) model. Multiply connected links, like the vertical along the
  left edge have much lower entropy than loop/dumbbell combinations,
  and hence give a relatively small contribution to the partition
  function.}
\end{figure}

This is a theory consisting of the field $\{\bvec{l}\}$ living on the
\emph{links} of the lattice, and the field $\{Q\}$ which lives on the
\emph{sites}. The field $\{Q\}$ is sibject the constraint
$\sum_{\bvec{x}} Q = 0$, i.e. overall charge neutrality, whereas the
field $\{\bvec{l}\}$  must satisfy the local constraint $\Delta \cdot
\bvec{l} = qQ$ on all lattice points.  The latter constraint means that every
$+Q/-Q$ pair must be joined by $q$ occupied links, in addition we can
have $\{\bvec{l}\}$ excitations which are not nucleated to any $+Q/-Q$
pairs, these must form \emph{closed} loops. \Figref{Z2Config} shows a
typical configuration for the $q=2$ model.

In the compact Abelian Higgs model considered in refs.
\onlinecite{Sudbo:2002,Smiseth:2003} the fields $\{\bvec{l}\}$ and
$\{Q\}$ represent \emph{vortices} and \emph{monopoles}, i.e. they are
the \emph{topological excitations} of the matter-field and gauge-field
respectively. That interpretation does \emph{not} apply in the current
case, but the interpretation of the $\{Q\}$ field is that it maintains
the discrete properties of the original theory \Eqref{PhaseZ_q}. With
$Q \equiv 0$ (the $q \to \infty$ limit), \Eqref{ZLoopGas} reduces to a
loop-gas with steric repulsion, this is a well known model with an
\emph{inverted XY} transition\cite{Dasgupta:1981}. Note that the
special case $q=1$ effectively represents no constraint. In this case,
the theory \Eqref{ZLoopGas} is noninteracting, and sustains no phase
transition.  For all $q \geq 2$ \Eqref{ZLoopGas} has a phase
transition, between a phase filled with link segments for $\beta >
\beta_c$, and a vacuum phase which does not contain link excitations.

We have performed Monte Carlo (MC) simulations of the $Z_q$ model,
using both a phase-representation, \Eqref{PhaseZ_q}, as well as the
loopgas/dumbbell gas (LDG) representation, \Eqref{ZLoopGas}. The phase
representation is simulated as a conventional spin simulation. In the
LDG representation, the fundamental Monte Carlo moves are represented
by alternating attempts of inserting a closed loop excitation of the
$\bvec{l}$ field and a dumbbell configuration consisting of a $+Q/-Q$
pair connected with an occupied $q$-valued link (the vertical link to
the left in \Figref{Z2Config} is an example of an elementary dumbbell
excitations). For $q=2$ the (vacuum) excitations of a loop or a
$+Q/-Q$ pair have the same energy, while for $q > 2$ the elementary
dumbbell excitations are more expensive than the elementary loop
excitations, and their relative importance diminishes with increasing
$q$.

\begin{figure}[htbp]
  \centerline{\scalebox{0.55}{\rotatebox{0.0}{\includegraphics{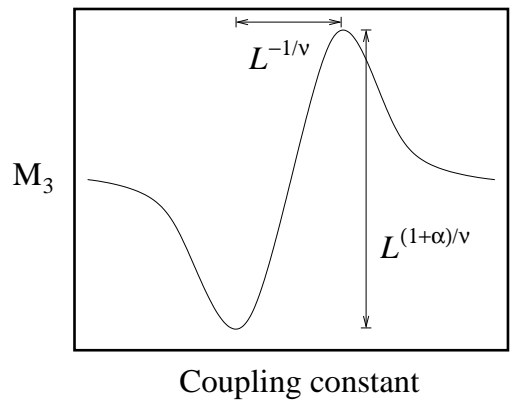}}}}
  \caption{\label{skjematisk}Schematic figure showing third moment of
    action, and how data are extracted for FSS analysis. For further
    details of this method see \protect\cite{Sudbo:2002}.}
\end{figure}

The main goal has been to determine how the critical properties change
with $q$. The central quantity we have considered is the connected third order
moment of the action \cite{Sudbo:2002}
\begin{equation}
  \label{STre}
  \langle (S - \langle S \rangle)^3 \rangle \propto \left| \beta - \beta_c \right|^{1 + \alpha},
\end{equation}
which recently has been demonstrated to yield surprisingly good scaling
results compared to second moments \cite{Sudbo:2002}.
When approaching the critical point, the correlation length $\xi$
diverges as $\xi \propto \left|\beta - \beta_c\right|^{-\nu}$.
Therefore, in a finite system of linear extent $L$ we find that 
the third order moment in \Eqref{STre} scales with $L$ as
\begin{equation}
  \label{STreL}
  \langle (S - \langle S \rangle)^3 \rangle \propto L^{\frac{1 + \alpha}{\nu}}.
\end{equation}
The main advantages of the third order moment in \Eqref{STre} are that
(1) good quality scaling is achieved for practical system sizes even
for models with $\alpha < 0$, e.g. the 3DXY model, and (2) one set of
measurements gives \emph{both} the combination $(1+ \alpha)/\nu$ and
$-1/\nu$ \emph{independently} \cite{Sudbo:2002}, although it is more
difficult to achieve high precision on the latter. A schematic figure
of $\langle \left(S - \langle S \rangle \right)^3\rangle$  as a
function of coupling constant is shown in \Figref{skjematisk}, and
figures \ref{MIIIzq} and \ref{MIIILG} show finite-size scaling (FSS)
of the peak to peak value.

\newcommand{\Bullet}{\text{\large $\bullet$}}
\newcommand{\Square}{\text{\tiny $\blacksquare$}}

\begin{figure}[htbp]
\centerline{\scalebox{0.35}{\rotatebox{-90.0}{\includegraphics{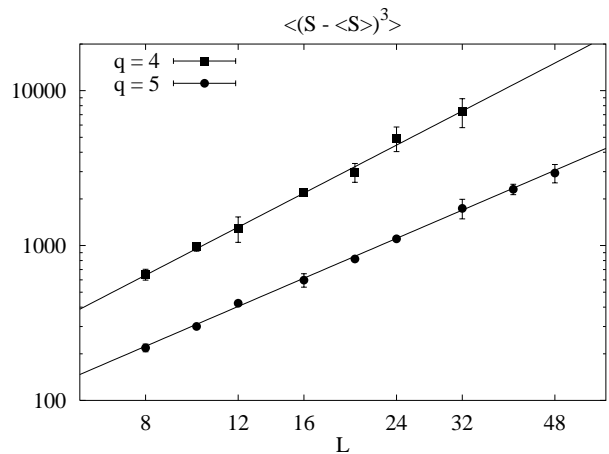}}}}
\caption{\label{MIIIzq}This figure shows the scaling of 
  $\langle\left(S - \langle S \rangle \right)^3\rangle$ for
  $q=4(\Square)$ and $q=5(\Bullet)$, the results are obtained using
  the phase representation \protect \Eqref{Z_qPart}. The $q=4$ results
  show $Z_2$ scaling with $(1+\alpha)/\nu = 1.76 \pm 0.05$, and the
  $q=5$ curve shows $XY$ scaling with $(1+ \alpha)/\nu = 1.46 \pm
  0.03$.}
\end{figure}

\begin{figure}[htbp]
\centerline{\scalebox{0.35}{\rotatebox{-90.0}{\includegraphics{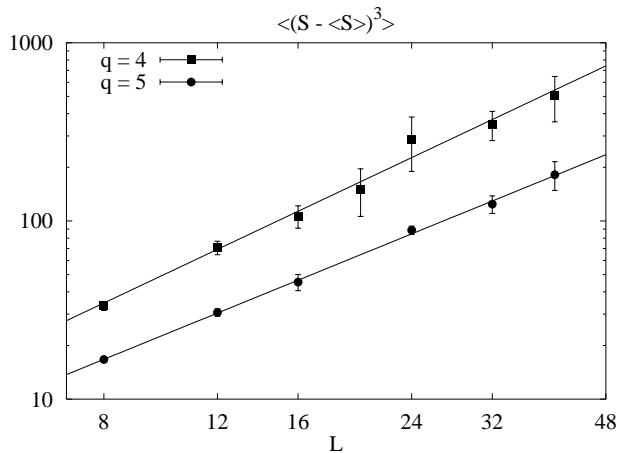}}}}
\caption{\label{MIIILG} This figure is similar to \protect
  \Figref{MIIIzq}, but the results are obtained using representation
  \Eqref{ZLoopGas}.  The $q=4$ results show $Z_2$ scaling with
  $(1+\alpha)/\nu = 1.70 \pm 0.05$, and the $q=5$ results scale with
  $(1 + \alpha)/\nu = 1.47 \pm 0.06$, i.e. qualitatively similar to
  the results in \protect\Figref{MIIIzq}}
\end{figure}

\begin{figure}[htbp]
\centerline{\scalebox{0.35}{\rotatebox{-90.0}{\includegraphics{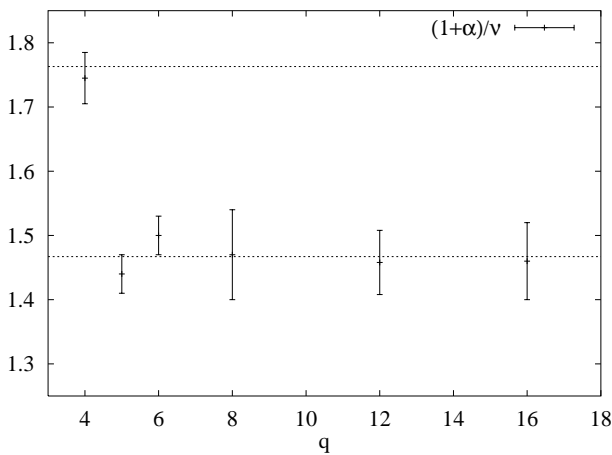}}}}
\caption{\label{univ_exp} The exponent-combination $(1+ \alpha)/\nu$ versus
  $q$. Note how it changes value abruptly as $q$ is increased from $q=4$ 
  to $q=5$. The dashed lines are the Ising $(Z
_2)$  and XY $(Z_\infty)$ values 
  of 1.763 and 1.467, respectively.}
\end{figure}

We have considered systems of size $L \times L \times L$ with
$L=8,10,12,16,20,24,32,40,48$, and up to $2 \cdot 10^7$ sweeps over
the lattice. In addition to the $q=4$ and $q=5$ presented in figures
\ref{MIIIzq} and \ref{MIIILG}, we have also studied the $q$ values
$q=6,8,12,16$ and $24$, ref. \onlinecite{Sudbo:2002} shows results of
$q=2$ simulations of \Eqref{ZLoopGas}. We find that the combination
$(1 + \alpha) / \nu$ changes abruptly from the $Z_2$ value of 1.763
\cite{Hasenbusch:1998} to the XY value of 1.467 \cite{Hasenbusch:2001}
when increasing $q$ from $q=4$ to $q=5$. A further increase of $q$
beyond $q=5$ does not affect the value of $(1 + \alpha)/\nu$, as shown
in \Figref{univ_exp}.

That the $Z_q$ model is in the XY universality class for $q \geq 5$
must imply that \emph{at the critical point} the discrete structure is
rendered irrelevant for these $q$ values. To investigate this point
further, we have implemented a simple real-space RG procedure, which
attempts to probe for what values of $q$ the discrete nature of $Z_q$
model is relevant at the critical point. We denote the untransformed
phases and fields as $\theta_0$. The renormalized phase at level $n+1$
is given by the \emph{block spin} construction
\begin{equation}
 \label{blockEq}
 \theta_{n+1}=\atan \left( \frac{\sum_{k} \sin \theta_n(k)}{\sum_{k} \cos \theta_n(k)} \right),
\end{equation}
where the sum over $k$ in \Eqref{blockEq} is over the eight spins in a
$2 \times 2 \times 2$ cube. For $q=2$, this transformation is clearly
trivial, since adding a number of phases $0$ and $\pi$ will still give $0$
or $\pi$. However, for $q>2$ the effective $q^{\ast}$ will increase
with $n$, and for $n\to \infty$ the resulting block spins can take
\emph{any} direction. 

\begin{figure}[htbp]
  \centerline{\scalebox{0.35}{\rotatebox{-90.0}{\includegraphics{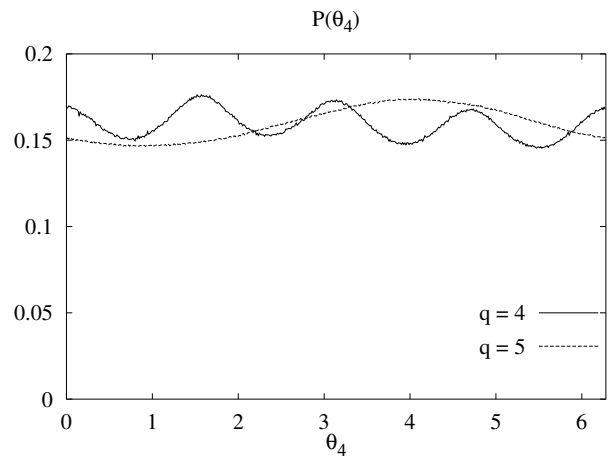}}}}
\caption{\label{Hist} Histograms of $\theta$ after four rescalings for
  of the critical state. For $q=4$, the distribution shows clear signs
  of a discrete background, whereas for $q=5$ this is not the
  case. The slow variation in the $q=5$ histogram is \emph{not}
  commensurable with a wavelength of $2\pi/5$, and probably only due
  to insufficent sampling.}
\end{figure}

We next investigate whether the system flows towards an infinite value of 
$q^{\ast}$ or not under such a RG transformation. This is tantamount to 
asking whether the discrete structure is rendered irrelevant or not on 
long length scales. To this end,  at  each iteration step  $n$, we have  
recorded \emph{histograms} $h_n(\theta)$ of the phase distributions
on the lattice, and monitored the manner in which this histogram flows 
under rescaling. By purely visual inspection we find that for $q=4$
the discrete nature of the $Z_q$ model persists, whereas for $q=5$ it
is washed away, this is illustrated in \Figref{Hist}.

\begin{figure}[htbp]
\centerline{\scalebox{0.45}{\rotatebox{-90.0}{\includegraphics{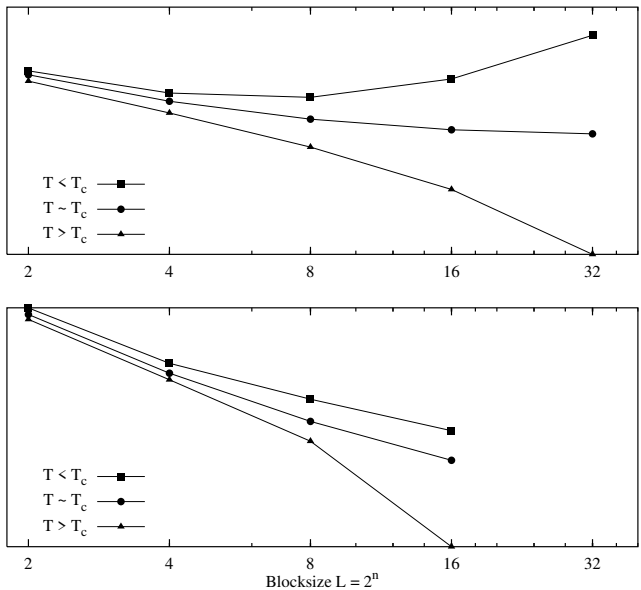}}}}
\caption{\label{FlowFig} The flow of the coefficient $a_{n,q}$ for $q=4$ and
  $q=5$. For $q=4$, we see that there is a fixed point at the critical
  point, whereas for $q=5$ we see that $a_{n,q}$ \emph{flows to zero
    at the critical point}. In the figure $a_{n,5}$ flows to zero also
  for $T < T_c$; this is a finite size effect. This coeffeicient will
  eventually flow to infinity for sufficiently large systems/low $T$.}
\end{figure}

To study this RG flow at a \emph{quantiative} level, we have written
the phase distribution $P_n(\theta_n)$ as a sum of harmonic functions
\begin{equation}
  \label{harmonic}
  P_n(\theta_n) = a_{n,0} + \sum_{k} \left(a_{n,k} \cos \left(\frac{k 2\pi}{q}\right) 
+ b_{n,k} \sin \left(\frac{k 2\pi}{q}\right) \right).
\end{equation}
Here, the coefficient $a_{n,k}$ in \Eqref{harmonic} denotes the $k$-th
Fourier-cosine component at RG level $n$. Clearly, the coefficient
$a_{n,q}$ is the interesting component, we have studied how this
coefficient flows under repeated rescaling. For $q=4$ this coefficent
shows critical fixed point behaviour, whereas for $q=5$ it flows to zero, even
for $T$ well below the critical temperature, this is shown in \Figref{FlowFig}.

Also the LDG representation \Eqref{ZLoopGas} gives a qualitative
indication that for $q \geq 5$ the discrete nature of the theory is
irrelevant. In this representation the discrete nature is represented
solely by the $Q$ excitations, so measurements of $\langle \left
  |Q\right| \rangle$ should give a quantitative indication of the the
importance of the discrete structure.  Measurements of $\langle |Q|
\rangle$ at the critical point give $\langle | Q | \rangle \approx
0.07, 5.9 \cdot 10^{-4}$ and $2.75 \cdot 10^{-6}$ for $q=2,4$ and $5$
respectively, whereas the link density $\langle \left| \bvec{l}
\right| \rangle \approx 0.15$ for all $q$. Hence at $q=5$ the discrete
$Q$ excitations hve been completely frozen out, and the tangle is
essentially identical to the \emph{pure-loop} tangle of the 3DXY
model.

In summary, we have determined the critical exponent combination $(1 +
\alpha)/\nu$ in the $d=3$ $Z_q$ spin model for $q \geq 4$.  Using two
different representations we have found that for $q \geq 5$, the
combination $(1 + \alpha)/\nu$ takes a value which is consistent with
the value taken in the $3DXY$ model. Along with other more qualitative
indicators this means that at the critical point discrete structure
finer than $q=5$ is irrelavant at the critical point, and the long
distance properties of the theory are determined by the larger
symmetry group $U(1)$. These results are in accordance with RG studies
starting with a $U(1)$ symmetric theory which is perturbed by a
perturbation with $Z_q$ symmetry.

We acknowledge support from the Norwegian Research Council through the
Norwegian High Performance Computing Centre (NOTUR). All computations
were carried out on an Origin SGI3800. Martin Hasenbusch is
acknowledged for valuable comments on an earlier version of the
manuscript.

\end{document}